\documentclass[twocolumn,twopage,a4paper]{article}
\usepackage[T1]{fontenc}
\usepackage[english]{babel}
\usepackage[dvips]{graphics}

\newcommand{\beq}{\begin{equation}}
\newcommand{\eeq}{\end{equation}}
\newcommand{\bma}{\begin{math}}
\newcommand{\ema}{\end{math}}
\newcommand{\beqa}{\begin{eqnarray}}
\newcommand{\eeqa}{\end{eqnarray}}

\def\expect#1{\langle\, #1\, \rangle}

\def\opone{\le\textbf{}\textbf{}avevmode\hbox{\small1\kern-3.8pt\normalsize1}}

\title{Density Matrix Renormalization Group Study of a Lowest Landau Level
Electron Gas on a Thin Cylinder}
\author{Emil J. Bergholtz and Anders Karlhede \\  \textsl{Department of Physics,
Stockholm University}\\
\textsl{AlbaNova University Center}\\ \textsl{SE-106 91 Stockholm,
Sweden}}
\begin{document}
\maketitle \textsl{We investigate the ground state properties of a
two-dimensional electron gas in the lowest Landau level using the
Density Matrix Renormalization Group. The electron gas is confined
to a cylinder with a strong magnetic field perpendicular to the
surface. For a thin cylinder, the ground state is for generic
filling factors a charge density wave state, however, at $\nu=1/2$
a homogeneous and apparently gapless ground state is found and
particle and hole excitations are studied.}

Of the many different methods used for studying strongly
correlated systems the Density Matrix Renormalization Group (DMRG)
invented by White \cite{White,White2} in 1992 is one of the most
successful. The method, which is an extension of Wilson's
Renormalization Group (RG) \cite{Wilson}, provides a reliable
method for analyzing one-dimensional quantum systems that are far
out of reach with exact diagonalization methods.  For higher
dimensional systems the DMRG is much less useful.

The two-dimensional electron gas in a perpendicular magnetic field
becomes  effectively  one-dimensional  when the field is strong
enough for the motion to be restricted to the lowest Landau level
(LLL).  In this limit the DMRG may  become an efficient tool. The
finite size DMRG has recently been applied to this system for an
electron gas confined to a torus \cite{Shibata}. The results are
in excellent agreement with exact diagonalization studies,
extending the latter from around 15 to 25 particles. In this
article, we apply the infinite size version of the DMRG to the
lowest Landau level problem when the electron gas is confined to a
cylinder.  This makes it possible to study the limit where the
number of particles $N$ tends to infinity. On the other hand, the
range of the interaction limits us to study thin cylinders.

For generic quantum Hall (QH) filling fractions such as $\nu=1/3$
we find charge density wave ground states where, for $\nu=1/3$,
approximately every third site is occupied by an electron. This
result is in agreement with previous results on smaller systems
\cite{Haldane} where the $1/3$ state is shown to be a charge
density wave on a thin cylinder. However, for the half-filled
Landau level we find a non-trivial result---a homogeneous ground
state  with excitations that appear to be gapless. (For even smaller
$L$, this state also gives way to a crystalline state.)

We consider a charged particle moving in two dimensions in a
perpendicular magnetic field $B$. By solving the  Schrödinger
equation in  Landau gauge, $\vec{A}=Bx\hat{y} $, we map the
two-dimensional system onto a one-dimensional system. The
eigenfunctions in the lowest Landau level are $\psi_{k}(\vec{r})=
\frac{1}{\sqrt{\pi^{1/2}L}}e^{iky}e^{-\frac{1}{2\ell^{2}}(x+k\ell^{2})^{2}}$,
where $L$ is the  circumference of the cylinder, $\ell
=\sqrt{\frac{\hbar c}{eB}}$ is the magnetic length and $k$ is the
momentum in the $y$-direction. Imposing periodic boundary
conditions in the $y$-direction gives $k=\frac{2\pi m}{L},  m \in
\mathbf{Z}$ and the density of states in the LLL is $n_{s}=1/2\pi
\ell^{2}$. The eigenfunction with $y$-momentum $k$ is centered at
$x=X_{k}=-k\ell^{2}$. This provides a mapping of the LLL onto a
one-dimensional problem which can be thought of as a spin chain
where each site $k$ has two states---either the site is occupied
by an electron or it is empty.

Following Trugman and Kivelson \cite{Kivelson}, we introduce a
repulsive short range potential between the particles which can be
expanded as
\begin{equation} V(\vec{r})=\sum_{j}c_{j}b^{2j}\nabla^{2j}\delta^{2}(\vec{r}),
  \label{generalint}\end{equation}
where $b$ is the range of the interaction. Taking the limit of
short range, $b\rightarrow 0$, the expression (\ref{generalint})
reduces to $V(\vec{r})=c_{1}b^{2}\nabla^{2}\delta^{2}(\vec{r})$
since $\expect{\delta^{2}(\vec{r})}=0$ for any antisymmetric
state. The form of this interaction can be motivated by the fact
that the physics of quantum Hall systems does not change
qualitatively if the interaction is made more local than the more
realistic Coulomb interaction.

From the interaction and the single particle states in the LLL we
find a second quantized form of the Hamiltonian. The Hamiltonian
is then analyzed and the low energy states are approximated by use
of the DMRG method. The Hamiltonian is
\begin{equation}
H=\mathcal{J}\sum_{n}\sum_{k>l}(k^{2}-l^{2})e^{-\frac{1}{2}(k^{2}+l^{2})}c_{l+n}^{\dag}c_{k+n}^{\dag}
c_{l+k+n}c_{n}, \label{AH} \end{equation} where $n$, $k$ and $l$
denote all possible momentum states and we use units where $\ell =
1$ and set $\mathcal{J}=1$. The DMRG algorithm becomes inefficient
when the interaction extends over too many sites. The most local
interaction in position space leads to the momentum space
interaction in (\ref{AH}). We introduce a cut-off $d$ in the range
of the interaction---$d$ is the number of sites the interaction
extends over. Since the distance between the sites is $2\pi
\ell^{2}/L$ it follows that we are restricted to small
circumferences $L$ in order for the cut-off not to modify the
shape of the interaction. Thus we are studying the QH system on a
{\it thin} cylinder. Note, however, that alternatively we may view
this as a large two-dimensional system with an interaction that
extends only a distance $\sim 2\pi d\ell^{2}/L$ in the
$x$-direction. Albeit being an unphysical interaction this is a
mathematically well-defined limit where the system can be studied.
We perform calculations for $d \leq 6$. From the form of the
interaction in (2) it follows that this restricts $L$ to at most
about 10 magnetic lengths. The  results presented below are for
$L=2\pi \ell$---this corresponds to the distance between the sites
being $\ell$.

As it stands, the Hamiltonian has a trivial ground state since the
interaction is repulsive -- the energy is minimal (and equal to
zero) when there are no particles present. However, in a real
physical system the density of electrons, $n_{e}$, is finite---the
negative charge of the electron gas is compensated for by a
positive background charge that makes the whole system neutral.
Thus we want to analyze the case when the filling fraction
$\nu\equiv \frac{n_{e}}{n_{s}}=2\pi \ell^{2} n_{e}$ is fixed and
nonzero. In order to target a state with  a specific filling
fraction, $\nu$, we add a term to the Hamiltonian that favours the
chosen density: $H'=H+\Delta H=H+ \gamma
N^{2}(\frac{1}{N}\sum_{i=1}^{N}c_{i}^{\dag}c_{i}-\nu)^{2}$. (All
energies presented below are of course for $\gamma=0$.)

The Hamiltonian  commutes with $\sum_{i}^{N}
c^{\dagger}_{i}c_{i}$, hence the total number of particles is a
conserved quantum number. As  a  consequence, for given $N$ only
the filling fractions $\nu=0,1/N,2/N,\ldots,(N-1)/N,1$ can be
obtained.  This makes it in general necessary to keep states with
different number of particles when studying a certain state in the
$N\rightarrow \infty$ limit.

\begin{figure}[h!]
\begin{center}
\resizebox{!}{55mm}{\includegraphics{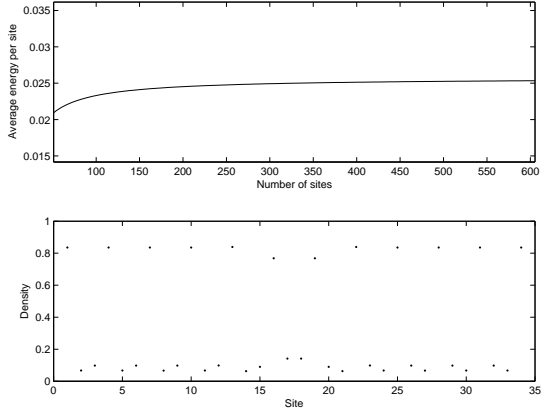}}
\end{center}\caption{\textit{{\small The average energy per site
      (above) and the density distribution
$\expect{c^{\dag}_{i}c_{i}}$ for 34 sites around the center of a
large system (below) for the $\nu=1/3$ ground state.}}}
\label{var033}
\end{figure}

For generic filling fractions we find charge density wave ground
states in agreement with earlier work \cite{Haldane,Tao}. As an
example, we show in Fig. \ref{var033} results for $\nu=1/3$. The
ground state is, to a good approximation, obtained by putting an
electron on every third site. Thus the range of the interaction is
so short that this simple crystal has lower energy than, e.g., a
homogeneous state \cite{Laughlin}.

However, for the half-filled  Landau level we find a very
different and  much more interesting result---a homogenous ground
state \cite{emil}.
\begin{figure}[h!]
\begin{center}
\resizebox{!}{55mm}{\includegraphics{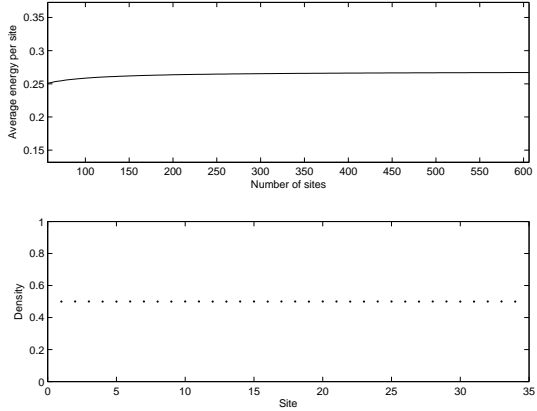}}
\end{center}\caption{\textit{{\small The $\nu=1/2$ ground state
energy per site (above) and the density distribution $\expect{c^{\dag}_{i}c_{i}}$
around the center of a large system (below). The
density distribution inside the system is totally uniform.
Furthermore, the energy converges smoothly. That the energy
increases is due to the fact that the edge effects become less
important when the system size increases.}}} \label{qh1}
\end{figure}
In Fig. \ref{qh1} we present some of our results  for the
$\nu=1/2$ state. The obtained ground state has an almost perfectly
uniform density throughout the system\footnote{This is true except
near the boundaries of the system. The repulsive interaction, in
combination with free boundary conditions, favours crystalline
states for small systems and also near the edges of large
systems.}. The mean density is $1/2$ up to the precision of at
least $16$ digits---this precision is a consequence of $N_{e}$
being a conserved quantum number. Furthermore, the standard
deviation of the single site density $\expect{c^{\dag}_{i}c_{i}}$
is as small as $10^{-12}$ for large systems. We also note that the
energy of the homogeneous state is very low compared to the
crystalline states obtained when $d \le 2$. Thus the obtained
state has used the off-diagonal elements of the Hamiltonian to
reduce the energy in a very efficient way. The results quoted are
independent of the number of states, $m$, that are kept in each
iteration.

The density $\expect{c^{\dag}_{i}c_{i}}$ given above, is really
the linear density along the cylinder, i.e. the full density
integrated around the cylinder. The density operator is
\begin{eqnarray}
\rho(\vec{r})=\sum_{m,n}\psi^{*}_{m}(\vec{r})\psi_{n}(\vec{r})c_{m}^{\dag}c_{n}
\nonumber\\\propto\sum_{m,n}e^{i(n-m)y}
e^{-\frac{1}{2}((x+n)^{2}+(x+m)^{2})}c_{m}^{\dag} c_{n}.
\label{density} \end{eqnarray} However, we find that
$\expect{c^{\dag}_{m}c_{n}}\propto\delta_{m,n}$ to a very high
precision\footnote{The deviation from this result is roughly of
the same order as the deviation of the single site density
$\expect{c^{\dag}_{i}c_{i}}$, i.e. $10^{-12}$.}. Hence the density
is constant on the cylinder\footnote{The finite distance $2\pi
  \ell^{2}/L=\ell$ between two sites leads to a variation in the
  spatial density of roughly $0.01\%$.}.

For $d \le 2$, the  $\nu=1/2$ ground state is crystalline. At
$d=1$ every other site is occupied and at $d=2$ two occupied sites
alternate with two empty---these are exact ground states. At $\nu
=1/4 $ the ground state is a crystal for $d\le 6$.

\begin{figure}[h]
\begin{center}
\resizebox{!}{55mm}{\includegraphics{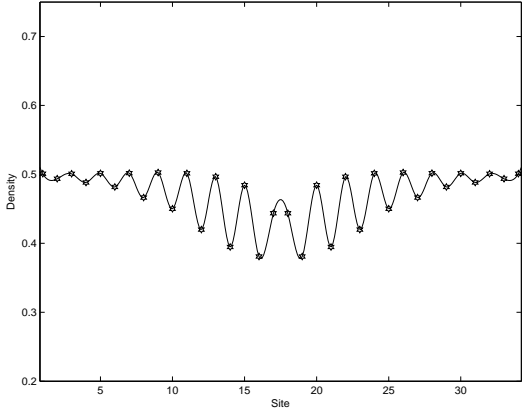}}
\end{center}\caption{\textit{{\small The density profile of a charge
one hole in the $\nu=1/2$ state. The plot shows the expectation
values of the single site density operators
$\expect{c^{\dag}_{i}c_{i}}$ for a range of sites near the center
of the superblock. The line is a guide to the eye.}}} \label{2}
\end{figure}

We have made  a preliminary investigation of  charged  excitations
at $\nu=1/2$ by targeting states with density $\nu=1/2 \pm 1/N$
for $N$ sites\footnote{When doing  this we target also the
$\nu=1/2$ state.}. We find a particle excitation with charge
$-|e|$ relative to the ground state and energy $E_{p}=2.0$  and a
hole with charge $|e|$ and energy $E_{h}=-1.8$  ($\mathcal{J}=1$,
$d=4$ and we keep $m=4$ states in total). This gives an energy gap
$\Delta=E_{p}+E_{h}=0.2$ to creating a separated particle-hole
pair. We take the smallness of $\Delta$ compared to $E_{p},E_{h}$
to indicate that $\Delta$ approaches zero in the thermodynamic
limit and  that the system is gapless. To test this we should
consider the limit $N \rightarrow \infty$ and $m \rightarrow
\infty$. For the particle this is straightforward, however, for
the hole the calculations are unstable  when $m$ is increased.
Increasing $N$ only moderately (to avoid the instability) we find
that $\Delta$ decreases with increasing $m$, supporting the
conclusion that $\Delta \rightarrow 0$. We believe  the
instability is due to a degeneracy in the energy of the hole
state. We find that the hole forms at the center of the
system---there should then be degenerate states where the hole is
translated to other positions. That no instability is seen for the
particle indicates that it forms at the edge of the system. Fig.
\ref{2} shows the density profile of the hole. The charge
integrated  over the area, with length $34\ell$, shown in the
figure is $0.98|e|$ relative to the ground state in Fig.
\ref{qh1}.

\begin{figure}[h]
\begin{center}
\resizebox{!}{55mm}{\includegraphics{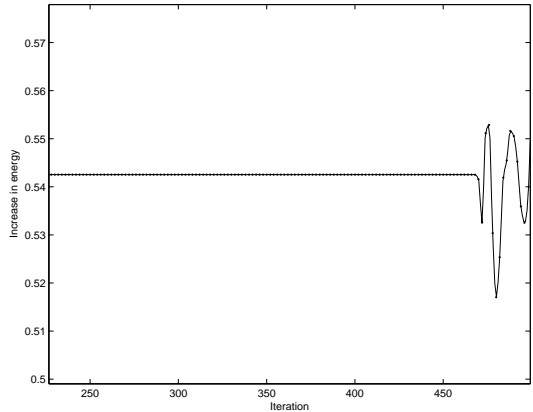}}
\end{center}\caption{\textit{{\small Energy added to ground state in each
iteration for half-filled LLL keeping states with  $\nu=1/2$ and
$\nu=1/2+1/N$. Initially, the ground state is homogeneous, then
eventually a transition to a state with a well separated
particle-hole pair is observed.}}} \label{new}
\end{figure}

A further indication of the existence of gapless excitations is
shown in Fig. \ref{new}. It shows the increase in total energy of
the $\nu=1/2$ ground state per iteration up to large
systems\footnote{The figure shows only data from every second
iteration since there is a ``parity-effect'' that distinguishes
even and odd iterations.}. In these calculations, states at density
$\nu=1/2$ {\it and} $\nu=1/2+1/N$ are kept. Initially, a smooth
convergence to the homogeneous $\nu=1/2$ ground state is
observed---this is reflected in the figure in that the same amount
of energy is added at each step. In each iteration two sites and
one electron are added at the center of the superblock, thus
keeping the average density fixed at $\nu=1/2$. Then suddenly at
iteration 470 a slightly smaller energy is added. A new ground
state is formed as a superposition of the homogeneous $\nu=1/2$
ground state in the previous iteration (adding two sites and one
electron to this) and the state containing  one particle
excitation (adding two sites and no particle to this). The energy
of this state is actually lower than the energy of the homogeneous
state since the added energy in the last iteration
decreased\footnote{Eventually, after  further iterations, the
state converges to a state with a well separated particle-hole
pair. That this is picked out rather than the homogeneous state
(or a linear combination) is a consequence of our algorithm.}. We
take this as indicating that the particle-hole excitation is
gapless.

In addition to the above studies of excitations, where we always
target only the lowest energy state, we have also targeted the two
or three lowest  states. We then find that these states become
degenerate  as $N$ and $m \rightarrow \infty$ thus lending further
support to the gaplessness of the ground state.

To conclude, for a thin cylinder we find a homogeneous, presumably
gapless, ground state for the half-filled lowest Landau level---at
all other filling fractions that we have investigated, including
$\nu=1/4$, we find charge density wave states. It is of course
tempting to identify the obtained ground state as the Rezayi-Read
state \cite{Rezayi}, that is known from comparison to exact
diagonalization studies to describe the $\nu=1/2$ state on the
sphere very well. It remains to compare our state to the
Rezayi-Read state.  It is computationally nontrivial to compare
large systems whereas small systems suffer from large edge
effects, presumably making a comparison meaningless. A natural
solution to this would be to employ the finite size DMRG method,
possibly on the torus. An interesting questions is why there is a
homogeneous state on thin cylinders only, as it seems, at $\nu =
1/2$. If this state is indeed the Rezayi-Read state, does our
results shed any further light on the $\nu = 1/2$ quantum Hall
state? A study of our $\nu = 1/2$ ground state indicates that it can
be understood in terms of free fermions \cite{ejbak}.

Anders Karlhede was supported by the Swedish Science Research
Council.

\end{document}